\documentclass[12pt]{amsart}
\usepackage{geometry}                
\usepackage{graphicx}
\usepackage{amssymb}
\usepackage{epstopdf}
\title{Flag Manifolds and Grassmannians}
\author{B. E. Eichinger}

\begin{document}
\maketitle

\begin{center}
{\small\itshape Department of Chemistry, University of Washington, Seattle, Washington 98195-1700}\par
\end{center}


\begin{abstract}
Flag manifolds are shown to describe the relations between configurations of distinguished points (topologically equivalent to punctures) embedded in a general spacetime manifold.  Grassmannians are flag manifolds with just two subsets of points selected out from a set of $N$ points.  The geometry of Grassmannians is determined by a group acting by linear fractional transformations, and the associated Lie algebra induces transitions between subspaces.  Curvature tensors are derived for a general flag manifold, showing that interactions between a subset of $k$ points and the remaining $N-k$ points in the configuration is determined by the coordinates in the flag manifold.  
\end{abstract}
\section*{Introduction}
The symmetric spaces known as Grassmannians have arisen in a wide variety of physical contexts in the last few decades. The simplest examples of Grassmannians are the projective spaces: $P\mathbb R^n$ with Fubini-Study metrics, K\"ahler manifolds, and hyper-K\"ahler manifolds.  Higher rank analogs have appeared in the computation of scattering amplitudes,\cite{Arkani} and they are associated with bipartite graphs \cite{Bipartite} and quivers.\cite{NekVaf}  This paper provides an overview of several aspects of these geometrical objects, highlighting a few features that have interesting physical consequences.   Flag manifolds over quaternions, which will be the primary focus, are related to non-commutative instantons.\cite{Nek}

\section*{Configuration Spaces}
A configuration space $M^N$, in the sense used here, will be a set of $N\ge 2$ points (topologically equivalent to punctures) embedded in an otherwise smooth spacetime manifold.   In representing physical particles the points are described by a module $\Psi_N(x)$, where $x$ is a set of coordinates to be determined.   The set of points will be divided into distinguishable subsets by defining a partition of $N$ into two or more integers $1\le k_\mu<N$ such that $\Sigma_\mu k_\mu = N$, with $k_\mu$ being the dimension of the $\mu$-th subset. To the $\mu$-th subset there is associated a module $\psi_\mu(x_N)$, with $\Psi_N=\bigoplus_\mu\psi_\mu$.  (The cardinality of the $\mu$-th subset will be understood to be $k_\mu$ to avoid multiple levels of subscripts in $\psi_\mu$.)  Each subset, taken individually, is assumed to have bounded measure such that the inner product $<\psi_\mu,\psi_\mu>$ is finite.  

A ring acts on $\Psi_N$ on the left.  The development described here can be done in the real or complex fields, or the quaternion ring. A symmetry group $U(k_\mu,\mathbb K)$ over a ring $\mathbb K = \mathbb {R,C,H}$ acting in some representation $h_\mu:\psi_\mu\to h_\mu\psi_\mu$ will preserve the inner product if $h_\mu$ is in $SO(k_\mu,\mathbb R)$, $SU(k_\mu,\mathbb C)$ or $U(k_\mu,\mathbb H)\sim Sp(k_\mu)$, depending on the chosen ring.  (To be precise, one should write this as a representation $\eta$ of $h_\mu$ acting by $\eta \phi_\mu$.  To conserve space and time this will usually be written simply as $h_\mu\psi_\mu$, but unless explicitly stated, this will mean the action of a selected but generic element of a representation of the group.  The number of points in the $\mu$-subset is subsumed in the dimension of the fundamental representation of the group $h_\mu$.  $\Psi_N$ conveys the properties of  the physical objects represented as points.)  The group $h_\mu$ is also expected to have an, as yet, unspecified action on coordinates.  This is simply denoted by $h_\mu:\psi_\mu(x_N)\to h_\mu\psi_\mu[x_N(h_\mu)]$ in anticipation of the action of a group on a representation.  One of our major objectives is to build an explicit set of coordinates and associated group action.  However, we are starting with a primitive notion of a four-dimensional spacetime manifold, and this suggests that the quaternion ring will be most useful for physical applications.  The interested reader will have no trouble translating the results to the real or complex field.  

Each subset, also called a system $A_\mu, 1\le\mu\le m$, comes equipped with a symmetry group $h_\mu\sim Sp(k_\mu)$.  The entire set of $N$ points, decomposed into the $m$ subsets, has a symmetry group acting by
\begin{equation}\label{Hrep}
H:\Psi_N(\cdot)\to \left[\begin{matrix} 
      h_1 & 0 &\cdot&0\\
      0 & h_2&0&\cdot\\
      \cdot&\cdot&\cdot&\cdot\\
      \cdot&\cdot&0&h_m\\
   \end{matrix}\right]\vee\left[\begin{matrix} 
      \psi_1 \\
      \psi_2\\
      \cdot\\
      \psi_m\\
   \end{matrix}\right].
 \end{equation}
Since $H=h_1\times h_2\times\cdots\times h_m=\bigotimes_\mu h_\mu$ and $\Psi_N$ is the direct sum of subspaces spanned by the $\psi_\mu$ as above, this is more simply written as $H:\Psi\to H\Psi$.  For now the group action on coordinates is acknowledged but not explicitly indicated.  The manner in which the group acts, as indicated by $\vee$, is also not specified.    

This construction leads to a set of subspaces that are independent of one another.  There is nothing in the structure of the $H$ that relates one subspace to another -- to do that one needs to enlarge the group for which the off-diagonal elements of 
\begin{equation}\label{Grep}
G:\Psi_N(\cdot)\to \left[\begin{matrix} 
      g_{11} & g_{12} &\cdot&g_{1m}\\
      g_{21} & g_{22}&g_{23}&\cdot\\
      \cdot&\cdot&\cdot&\cdot\\
      \cdot&\cdot&g_{m,m-1}&g_{mm}\\
   \end{matrix}\right]\vee\left[\begin{matrix} 
      \psi_1 \\
      \psi_2\\
      \cdot\\
      \psi_m\\
   \end{matrix}\right]
 \end{equation}
couple the $\psi_\mu$ subspaces to one another.   Since each subspace has a finite inner product, so too does $\Psi_N$ under the action of a compact group $G$.  The homogeneity of the subspace decomposition, implicit in the notion that $N$ can be partitioned in many different ways, recommends that $G$ and $H$ belong to the same group, which we now take to be $G\sim Sp(N)$.\cite{Helg}  Each matrix element in $Sp(N)$ is a quaternion.  $H$ is a maximal subgroup embedded in $G$, and this leads to consideration of the natural principal bundle, $G(G/H,H)$.\cite{K&N}  The subspaces interact with one another under the action of $Sp(N)$, signifying that the coset $G/H$ induces changes in the $\psi_\mu$ owing to the presence of other subspaces.  The coset space $G/H\sim Sp(N)/\bigotimes Sp(k_\mu)$ is a flag manifold.\cite{Mare}  A hyper-K\"ahler manifold corresponds to the partition $\{k_1=1, k_2=N-1\}$.

\section*{Grassmannians}
The simplest examples of flag manifolds are those for which the $N$-point configuration space is divided into just two subspaces.  In our case the associated symmetry group $H$ is the product of just two subgroups, $H=h_k\times h_n=Sp(k)\times Sp(n)$ with $k+n=N$ and $k\le n$. The coset space $Sp(k+n)/Sp(k)\times Sp(n)\sim G/H$ is a Grassmann manifold. Note that the notation has been changed so that the subscript $j$ on $\mathfrak h_j$ is the dimension of the fundamental representation of the group.  In the Grassmannian context this notation is more convenient than that used for the general $m$-component flag manifold.  

The Lie algebra $\mathfrak{sp}(j)$ of the group $Sp(j)$ consists of skew-symmetric matrices over the quaternion ring: $\mathfrak x^*=-\mathfrak x; \mathfrak x\in \mathfrak{sp}(j)$, where $\mathfrak x^*$ is the conjugate transpose of $\mathfrak x$.\cite{Helg} The real dimension of $Sp(j)$ is $j(2j+1)$.  The algebra $\mathfrak h_k\oplus \mathfrak h_n\in \mathfrak{sp}(k)\oplus \mathfrak{sp}(n)$, with $\mathfrak h^*_j=-\mathfrak h_j$,  consists of diagonal blocks sitting inside $\mathfrak x\in \mathfrak{sp}(n+k)$:
\[
\mathfrak x=\left[\begin{matrix} 
      \mathfrak h_k & \mathfrak p \\
      -\mathfrak p^* & \mathfrak h_n\\
   \end{matrix}\right],
\]
where $\mathfrak p$ is a $k\times n$ matrix with quaternion entries.  This gives a parameterization of $x\in Sp(k+n)$ as
 \[
x=\exp\left[\begin{matrix} 
     0 & \mathfrak p \\
      -\mathfrak p^* & 0\\
   \end{matrix}\right]\exp\left[\begin{matrix} 
      \mathfrak h_k & 0 \\
      0 & \mathfrak h_n\\
   \end{matrix}\right]=yh
\]
where the coset $y\sim G/H$ comprises the Grassmann manifold: it is a cross-section of the bundle with fiber H.  The elements of the coset space $y$ couple the individual components of $\psi_k$ to those of $\psi_n$, which can be encoded in a bipartite graph.\cite{Bipartite}  

A coordinate version of the Grassmannian will be helpful to expose several aspects of the geometry.  The Stiefel manifold $X=[X_k,X_n]\subset g$ comprises, say, the first $k$ rows of a matrix $g\in Sp(k+n)$.  Here $X_k$ is a $k\times k$ matrix and $X_n$ is a $k\times n$ matrix such that $XX^*=1$. (Here and throughout the dimension of the unit matrix $1$ will be understood from the context.)  The $X_k$ part of $X$ can be factored: 
\begin{equation*}\label{proj}
XX^*=X_kX^*_k+X_nX_n^*=X_k(1+YY^*)X^*_k=1
\end{equation*}
with $Y=X^{-1}_kX_n$, from which it follows that $1+YY^*=(X^*_kX_k)^{-1}\ge 1$.  The projective space $Y$ is the Grassmann manifold.   The real dimension of $X$ is $4k(k+n)-k-4k(k-1)/2=4kn+2k^2+k$. the dimension of $Y$ is $4kn$, and the remaining part, $X^*_kX_k$ has real dimension $k(2k+1)$.

Since $Xg_1$, $g_1\in Sp(k+n)$, is also a subset of  $Sp(k+n)$, it follows that $g_1$ acts on $X$ by 
\begin{equation*}
g_1:X\to \hat X=
[X_k,X_n]\left[{\begin{matrix}
A^*& -C^* \\
-B^* &D^*\\
\end{matrix}}\right]=[X_k A^*-X_n B^*, -X_k C^*+X_n D^*], 
\end{equation*}
where the partitioning of $g_1$ is compatible with that of $X$.  (The reason for the peculiar labeling of matrix elements will become apparent shortly.)  The transformation $X\to \hat X$ acts by a linear fractional transformation on the Grassmannian, sending $Y\to \hat Y$, such that 
\begin{equation*}
g_1:Y \to \hat Y =(A^*-YB^*)^{-1}(-C^*+YD^*).
\end{equation*}
Given that $g_1g_1^*=1$, it follows that 
\begin{equation}\label{map}
\hat Y= (A^*-Y B^*)^{-1}(-C^*+Y D^*) = (AY+B)(CY + D)^{-1}.
\end{equation}
The right hand version of this equation is canonical in the literature,\cite{Hua} which justifies the choice made for the matrix elements in $g_1$.   Linear fractional transformations are composed of rotations, translations, and inversions.

Now fix $g_1$ and vary $Y$; using eq. (\ref{map}) and the unitarity (over $\mathbb H$) of the group it is not difficult to show that 
\begin{align*}
d\hat Y=&{}(A^*-Y B^*)^{-1}dY(CY + D)^{-1},\\
1+\hat Y \hat Y^*=&{}(A^*-Y B^*)^{-1}(1+Y Y^*)(A-BY^*)^{-1},\textrm {and}\\
1+\hat Y^*\hat Y=&{}( Y^*C^* + D^*)^{-1}(1+Y^* Y)(CY + D)^{-1}.
\end{align*}
These pieces are assembled to give the invariant metric
\begin{equation*}\label{metric}
ds^2=\text{Tr}[(1+YY^*)^{-1}dY(1+Y^*Y)^{-1}dY^*].
\end{equation*}
In eq. (\ref{map}) one sees that $Y=0$ is mapped to $\hat Y=BD^{-1}=-(A^*)^{-1}C^*$ , which provides an alternative and very useful representation of the Grassmannian.  This identification gives a mixed expression for the metric:
\begin{equation*}\label{metric}
ds^2=\text{Tr}[(AA^*)dY(DD^*)dY^*]=\text{Tr}[(A^*dYD)(A^*dYD)^*].
\end{equation*}
However, $dY=(dB-YdD)D^{-1}=[dB+(A^*)^{-1}C^*dD]D^{-1}=(A^*)^{-1}(A^*dB+C^*dD)D^{-1}$.  The last version expresses the one-form $dY$ in terms of the off-diagonal element of the connection form
\begin{align*}
\omega=g^*dg=&{}\left[{\begin{matrix}
A^*& -C^* \\
-B^* &D^*\\
\end{matrix}}\right]\left[{\begin{matrix}
dA&-dB \\
-dC &dD\\
\end{matrix}}\right]=\left[{\begin{matrix}
A^*dA+C^*dC& -A^*dB-C^*dD \\
-B^*dA-D^*dC &B^*dB+D^*dD\\
\end{matrix}}\right]\\=&{}\left[{\begin{matrix}
\omega_{11}& \omega_{12}\\
-\omega^*_{12} &\omega_{22}\\
\end{matrix}}\right],
\end{align*}
giving 
\begin{equation}\label{w12}
A^*dYD=-\omega_{12}
\end{equation}
and 
\[
ds^2=\text{Tr}(\omega_{12}\omega^*_{12})
\]
for the coordinate-free version of the metric.

The connection form $\omega$ is conveniently constructed on the cotangent bundle.  The cotangent space has a basis ${\bf e}=({\bf e_1},{\bf e}_2,\cdots,{\bf e}_N)$ on which $G$ acts to the right.  Let ${\bf e}^0$ denote the basis at the identity of $G$, corresponding to the point $p^0\in x$.  The basis ${\bf e}(p)$ at an arbitrary point $p$ can be pulled back to the identity by the action of $g^{-1}\in G$.  That is, ${\bf e}(p)={\bf e}^0g$.  The change in the basis in an infinitesmal neighborhood of $p$ is $d{\bf e}={\bf e}^0dg={\bf e}g^{-1}dg$, and for our group this is\cite{Chern95}
\begin{equation*}\label{de}
d{\bf e}={\bf e}g^*dg={\bf e}\omega, 
\end{equation*}  
showing that $\omega$ is the connection form on the cotangent bundle of the configuration space $M^N$.

A $k\times n$ Grassmannian connects the points (vertices or nodes) in two different subsets, corresponding to a bipartite graph of $k$ white nodes and $n$ black nodes, with all possible connections constituting the $4kn$ real dimensions of the quaternionic Grassmannian.  (The connections within the $\mu$-th subset are contained in $h_\mu$ in a particular way, as will be seen later.)  Within a given bipartite graph one may select subsets of vertices from both the white and black nodes, which implies that Schubert varieties will be of interest.  The corresponding graphs will have external edges which connect the selected nodes to those remaining in the $(k+n)$-vertex set.  However, the full $(k+n)$-particle system has no connections to anything else by construction.  To make those connections the system must be expanded to include a third subspace, thought of as the surroundings.  (Connections to the surroundings will require a non-trivial but practical truncation scheme for calculations.)  The next section provides one aspect of the connections between subsets.

\subsection*{Curvature Tensors on Flag Manifolds}
The exterior derivative of $\omega$ is $d\omega =d(g^*dg)=dg^*\wedge dg = -g^*dgg^*\wedge dg=-g^*dg\wedge g^* dg=-\omega\wedge\omega$, giving 
\begin{equation}\label{MC2}
d\omega+\omega\wedge\omega =0.
\end{equation}
This is the second Maurer-Cartan equation (MC2).\cite{Cartan} It implies that the affine group acts on the tangent space of  the configuration space, \emph{i.e.}, the tangent and cotangent bundles are horizontal.\cite{K&N}

The Maurer-Cartan two-form fits into the setting of the flag manifold as follows.  Corresponding to the $\{k_\mu\}$ partition of $N$, denote the blocks of $g\in Sp(N)$ by $g_{\alpha\beta}$.  Similarly, $\omega = (\omega_{\alpha\beta}), 1\le\alpha,\beta\le m$ is the matrix written in block form.  With this partitioning in mind, eq. (\ref{MC2}) becomes  
\begin{equation*}
d\omega_{\mu\nu}+\Sigma^m_{\alpha=1}\omega_{\mu\alpha}\wedge \omega_{\alpha\nu}=0
\end{equation*}
and in particular, a block on the diagonal is
\begin{equation}\label{MCB}
d\omega_{\mu\mu}+\omega_{\mu\mu}\wedge\omega_{\mu\mu}+\Sigma_{\alpha\ne \mu}\omega_{\mu\alpha}\wedge \omega_{\alpha\mu}=0.
\end{equation}
(Note that the summation convention is not used.) 

Now suppose that $g\to H$. The corresponding connection form reduces: $\omega\to\bar\omega=\bar\omega_{11}\times\bar\omega_{22}\times\cdots\times\bar\omega_{mm}$.  All off-diagonal blocks of $\omega$ vanish, and one is left with $d\bar\omega_{\mu\mu}+\bar\omega_{\mu\mu}\wedge\bar\omega_{\mu\mu}=0$ for all $\mu$.  The tangent bundles for the $m$-subspaces are all horizontal, just as is tangent bundle for $M^N$.  Thus the tangent bundles are independent of one another -- all subspaces with vanishing MC2 equations are independent of the other subspaces.  

However, in a general flag manifold the off-diagonal elements do not vanish, and when $d\omega+\omega\wedge\omega\ne 0$, Cartan\cite{Cartan} identifies the obstruction as the curvature two-form $\Omega$, \emph{i.e.}, $d\omega+\omega\wedge\omega=\Omega$.  The diagonal part, $d\omega_{\mu\mu}+\omega_{\mu\mu}\wedge\omega_{\mu\mu}$ in eq. (\ref{MCB}), is horizontal and the off-diagonal part, which couples the basis vectors that are orthogonal to the $k_\mu$-subspace, is vertical.\cite{K&N}  It follows from eq. (\ref{MCB}) that a curvature two-form,\cite{Chern46}
\begin{equation}\label{curv}
\Omega_{\mu}=- \Sigma_{\alpha\ne \mu}\omega_{\mu\alpha}\wedge \omega_{\alpha\mu}=\Sigma_{\alpha\ne \mu}\omega_{\mu\alpha}\wedge \omega_{\mu\alpha}^*,
\end{equation}
is associated to every subspace in an irreducible representation of $Sp(N)$.  The skew-symmetry of $\omega$ was used to get the second equality.  The cuvature two-forms are clearly non-negative definite. Chern classes may be constructed from the two-forms:
\[
(\wedge\Omega_\mu)^\ell=\textrm{tr}(\Omega_\mu\wedge\Omega_\mu\wedge\cdots\wedge\Omega_\mu),\quad \ell\;\textrm{terms}.
\]

For the given $k_\mu$-partitioning, consider a change of basis $\hat{\bf e} = {\bf e}h$, which corresponds to a different selection of cross-section of $G/H$ ($h\subset H$).\cite{Chern95}  The associated connection form is defined by $d\hat{\bf e}=\hat{\bf e}\hat\omega$.  The exterior derivative of $\hat{\bf e}$ is 
\begin{equation*}
d\hat{\bf e}=d{\bf e}h +{\bf e}dh={\bf e}\omega h+{\bf e}dh={\bf e}h\hat\omega=\hat{\bf e}\hat\omega
\end{equation*}
giving $h\hat\omega=\omega h+dh$.  The exterior derivative of this equation gives 
\begin {equation*}\label{chg}
\hat\Omega=d\hat\omega+\hat\omega\wedge\hat\omega = h^*(d\omega+\omega\wedge\omega) h=h^*\Omega h
\end{equation*}
demonstrating the tensor character of $\Omega$.  This can also be an aid in simplifying the $\Omega_{\mu\mu}=\Sigma_{\nu\ne\mu}\omega_{\mu\nu}\wedge\omega^*_{\mu\nu}$ forms.

The off-diagonal components of the connection form are also of interest, and will help to shed light on the interpretation of the curvature two-forms.  To make this point it is convenient to consider just two subspaces, \emph{i.e.}, the Grassmann case.  The elements of the exterior derivative of the $\omega_{12}$ block in eq. (\ref{MC2}) is 
\[
d\omega_{12}=- \omega_{11}\wedge\omega_{12}-\omega_{12}\wedge\omega_{22},
\]
The exterior derivative of this equation gives
\begin{equation*}\label{eq*}
\Omega_{1}\wedge \omega_{12}=\omega_{12}\wedge\Omega_{2}
\end{equation*}
which is equivalent to vanishing torsion.  It also serves to show that the magnitudes of the two curvature tensors are equal in their projections onto the vertical component of the connection form. 

Continuing with the case of two subspaces, it follows that from eq. (\ref{w12}) that
\begin{align*}\label{Omega1}
\Omega_{1}=&{}A^*dYD\wedge D^* dY^*A=A^*dY(1+Y^*Y)^{-1}\wedge dY^*A\\
\Omega_{2}=&{}D^*dY^*A\wedge A^* dYD=D^*dY^*(1+YY^*)^{-1}\wedge dYD.
\end{align*}  
for which the obvious traces are 
\begin{align*}
R_1=&{}\textrm{tr}[(1+YY^*)^{-1}dY(1+Y^*Y)^{-1}\wedge dY^*]\\
R_2=&{}\textrm{tr}[(1+Y^*Y)^{-1}dY^*(1+YY^*)^{-1}\wedge dY].
\end{align*}  
These are respectively the Ricci two-forms associated with the two subspaces.  The components of the Ricci two-forms are those of the metric tensor, which identifies the Grassmannian as an Einstein space.\cite{Besse}

The physical implication of all this is that each subspace has an associated curvature two-form that is determined by the interactions with all of the other subspaces.  Where $M^N$ has been subdivided into just two subspaces, which might be thought of as a system and its surroundings, the magnitudes (eigenvalues) of the two-forms are equal, suggestive of Newton's Third Law. 
 
\section*{Yang-Mills Geometry}
The curvature forms specialize to the case of just two points, for which $\omega_{12}$ is a single quaternion: $Y\to q$.  The curvature two-forms may be rotated by elements of $Sp(1)$ bringing $A$ and $D$ to forms that commute with the identity. This yields 
\begin{equation*}
\Omega_1=(1+q\bar q)^{-2}dq\wedge d\bar q \quad {\rm and}\quad \Omega_2=\bar\Omega_1=(1+q\bar q)^{-2}d\bar q\wedge dq.
\end{equation*}
These are the curvature two-forms for the original Yang-Mills (YM) theory.\cite{YM,Belavin,At} This demonstrates that $Sp(2)/Sp(1)\times Sp(1)$ is the underlying YM geometry.  As Atiyah\cite{At} points out, $dq\wedge d\bar q$ is self-dual and $d\bar q\wedge dq$ is anti-self-dual.  One goes into the other by reflection of the 3-space (conjugation).

Using well-known group isomorphisms, $Sp(2)/Sp(1)\times Sp(1)\sim SO(5)/SO(3)\times SO(4)\sim SO(5)/SO(4)\sim S^4$,\cite{Helg,Lawson} the original Yang-Mills construction yields a 4-sphere geometry.  This suggests that one interpret the two $Sp(1)$ factors as the symmetry groups of two spins, located at antipodes of the sphere.  The only available coordinates in this construction are those defining the sphere, \emph{i.e.}, the coordinates of the Grassmannian.  The general flag manifold $Sp(N)/Sp(1)^N$ constitutes a many-body Yang-Mills theory.  The symmetry group of each of the $\psi_\mu(k_\mu=1)$ is $Sp(1)$, corresponding to the spin degree of freedom of the fiber sitting on each function.
\section*{Lie Algebra}
The infinitesmal generators of the Lie algebra for the symplectic group are built from the Grassmann coordinates, and this is best done in the $M(2,\mathbb C)$ basis for quaternions. Furthermore, it is convenient to use different symbols to label row and column indices of the $\mathbb H$-valued $k\times n$ Grassmannian.  Define $Q = (\zeta_{\alpha a}); 1\le \alpha \le 2k, 1\le a \le 2n$ to be the Grassmannian matrix of quaternions in the $M(2,\mathbb C)$ representation, with 
\begin{equation}\label{M2C}
\left[\begin{matrix} 
    \zeta_{2\mu - 1,2t-1} & \zeta_{2\mu - 1,2t} \\
     \zeta_{2\mu ,2t-1} & \zeta_{2\mu,2t}\\
   \end{matrix}\right]
=\left[\begin{matrix} 
      z^{(1)} & z^{(2)} \\
      -\bar z^{(2)} & \bar z^{(1)}\\
   \end{matrix}\right]_{\mu t}=q_{\mu t}; 1\le\mu\le k, 1\le t \le n.
\end{equation}
In this representation, $z^{(1)}_{\mu t}=x_{0,\mu t}+ix_{3,\mu t}$ and $z^{(2)}_{\mu t}=x_{1,\mu t}+ix_{2,\mu t}$, with $\bar z$ being the complex conjugate of $z$.  The differential operator $\partial/\partial \zeta_{\alpha a}$ is now defined such that $\partial \zeta_{\alpha a}/\partial \zeta_{\beta b}=\partial_{\beta b}\zeta_{\alpha a} =\delta_{\alpha\beta}\delta_{ab}$.  As an operator in $SU(2)$ this is
\[
\partial_{\mu t}=\left[\begin{matrix}
\partial/\partial z^{(1)} & \partial/\partial z^{(2)}\\
-\partial/\partial \bar{z}^{(2)} & \partial/\partial \bar{z}^{(1)}\\
\end{matrix}\right]_{\mu t}.
\]

There is an additional `almost complex' structure in the symplectic group that is helpful in computations.  One of the $su(2)$ basis vectors is 
\[
j=\left[\begin{matrix} 
   0 & 1 \\
   -1 & 0\\
   \end{matrix}\right], 
\]
and its action on a quaternion is complex conjugation:  
\[
j'q j=\left[\begin{matrix} 
   0 & -1 \\
   1 & 0\\
   \end{matrix}\right]\left[\begin{matrix}
      z_1 & z_2 \\
      -\bar z_2 & \bar z_1\\
   \end{matrix}\right]\left[\begin{matrix} 
   0 & 1 \\
   -1 & 0\\
   \end{matrix}\right]=\left[\begin{matrix} 
      \bar z_1 & \bar z_2 \\
      - z_2 & z_1\\
   \end{matrix}\right]=\bar {q}.
\]
The quaternion conjugate to $q$ is $q^*=j'q'j$ in the $M(2,\mathbb C)$ basis.  The skew-symmetry of the Lie algebra of $Sp(k+n)/Sp(k)\times Sp(n)$ requires elements $-w^*$ conjugate to $w$, and these comprise the matrix $-Q^*=-\bar {Q}'$.

The utility of the preceding representation of conjugation is that it facilitates differentiation of conjugate quaternions (the summation convention is now used): 
\[
\partial \bar \zeta_{\alpha a}/\partial \zeta_{\beta b}=\partial_{\beta b}\bar\zeta_{\alpha a}=\partial_{\beta b}(J'_{\alpha \gamma}\zeta_{\gamma c}J_{ca})=J'_{\alpha\beta}J'_{ab}=J_{\beta\alpha}J_{ba}.  
\]
In the following the short-hand notation $J'QJ\to \bar Q$ will be used: it is understood that the pre- and post-$J$ factors are of the form $1\otimes j$ (direct product) with appropriate dimension of the unit matrix, $1$, to be compatible with the $2k\times 2n$ matrix $Q$.  Now it is seen why use of two different symbols for row and column are useful -- it keeps the pre- and post-multiplicative $J$ factors straight.

The infinitesimal generators of the Lie algebra of $Sp(n+k)$ are parameterized by the coordinates of the Grassmannian, stated here without proof (the summation convention is used, and note also that to avoid a plethora of subscripts, $h$ and $H$ correspond to $h_1$ and $h_2$, respectively):
\begin{align*}
h_{\alpha \beta}&=  \zeta_{\alpha b}\partial_{\beta b}-\bar \zeta_{\beta b}\bar\partial_{\alpha b}\\ 
H_{ab} & = \zeta_{\mu a}\partial_{\mu b}-\bar \zeta_{\mu b}\bar\partial_{\mu a}\\
p_{\alpha a}&=   \bar\partial_{\alpha a}+ \zeta_{\alpha b}\zeta_{\mu a}\partial_{\mu b}\\
&=(\delta_{\alpha\beta}+\zeta_{\alpha b}\bar \zeta_{\beta b})\bar \partial_{\beta a}+\zeta_{\alpha b}H_{ab}\\
&=(\delta_{ab}+\zeta_{\mu a}\bar \zeta_{\mu b})\bar \partial_{\alpha b}+\zeta_{\mu a}h_{\alpha\mu}.
\end{align*}
It is easy to see that $h^{*}=-h, h\sim sp(k)$ and $H^{*}=-H, H\sim sp(n)$.  Furthermore, $\bar p$ differs from $p$ by quaternion conjugation as shown above.  The infinitesimal generators are written more succinctly as 
\begin{align*}
h=&{}Q\partial'-(Q\partial')^\ast\\
H=&{}Q'\partial-(Q'\partial)^\ast\\
p=&{}(1+QQ^\ast)\bar\partial+QH'.\\
\end{align*}

There are many different ways of writing these equations.  Let $r_{\alpha\beta}=\zeta_{\alpha a}\partial_{\beta a}$, so that $h=r-r^*=r-\bar r'=r-J'r'J$.  It follows that $hJ=rJ+(rJ)'$, which is clearly symmetrical: in components $(hJ)_{\alpha\beta}=(rJ)_{\alpha\beta}+(rJ)_{\beta\alpha}$.  By encapsulating the generators in these symmetrical forms the commutators are more symmetrical than they would be otherwise.  They satisfy the following commutation relations:
\begin{align*}
[(hJ)_{\alpha \beta},(hJ)_{\mu \nu}]={}&-J_{\alpha\mu}(hJ)_{\beta\nu}-J_{\alpha\nu}(hJ)_{\beta\mu}-J_{\beta\mu}(hJ)_{\alpha\nu}-J_{\beta\nu}(hJ)_{\alpha\mu}\\
[(HJ)_{ab},(HJ)_{cd}]={}&-J_{ac}(HJ)_{bd}-J_{ad}(HJ)_{bc}-J_{bc}(HJ)_{ad}-J_{bd}(HJ)_{ac}\\
[h_{\alpha \beta},H_{ab}]={}&0\\
\dag\quad\quad\quad[(hJ)_{\mu\nu},p_{\alpha a}]={}&J_{\alpha\mu}p_{\nu a}+J_{\alpha \nu }p_{\mu a}\\
\ddag\quad\quad\quad[(HJ)_{bc},p_{\alpha a}]={}&J_{ab}p_{\alpha c}+J_{ac}p_{\alpha b}\\
[p_{\alpha a},p_{\beta b}]={}&-J_{ab}(hJ)_{\alpha \beta}-J_{\alpha \beta }(HJ)_{ab}\\
[\bar p_{\alpha a},p_{\beta b}]={}&\delta_{\alpha \beta}H_{ba}+\delta_{ab}h_{\beta \alpha}
\end{align*}
The last two of these equations show that $\mathfrak{p}$ generates the entire $\mathfrak{sp}(k+n)$ Lie algebra as it must.

The upshot of the commutation relations is the following.  From eqs. (\dag) and (\ddag) it follows that 
 \begin{align*}
 [h_{\mu\nu},p_{\alpha a}]=&{}\delta_{\alpha\nu}p_{\mu a}+J_{\alpha\mu}(Jp)_{\nu a}\\
 [H_{bc},\bar p_{\alpha a}]=&{}-\delta_{ab}\bar p_{\alpha c}+J_{ac}(\bar pJ)_{\alpha b}.
 \end{align*}
 Alternatively,
 \begin{align*}
 [h_{\mu\nu},\bar p_{\alpha a}]=&{}-\delta_{\alpha\mu}\bar p_{\nu a}-J_{\alpha\nu}(J\bar p)_{\mu a}\\
 [H_{bc},p_{\alpha a}]=&{}\delta_{ac}p_{\alpha b}-J_{ab}(pJ)_{\alpha c},
 \end{align*}
where the versions with $\bar p_{\alpha a}$ are obtained from eqs. (\dag) and (\ddag) by complex conjugation and using the skew-symmetry of $h$ and $H$.  

Given two functions, $v$ in the representation space associated with subset $A_1$ and $V$ in that for subset $A_2$, the diagonal elements of the operators act by $h_{\mu\mu}v_\mu(n_\mu)=n_\mu v_\mu(n_\mu)$ and $H_{bb}V_b(n_b)=n_bV_b(n_b)$ (the summation convention is suspended in this paragraph).  Then $p_{\mu b}$ acts on $v_\mu$ as a raising operator, while $\bar p_{\mu b}$ acting on $V_b$ is a lowering operator.  Switching $p_{\mu b}$ and $\bar p_{\mu b}$ gives the opposite action.  The  presence of raising and lowering operators is expected, but the important aspect of the $\mathfrak p$ generators is that they act between different subspaces, and in so doing transfer excitations between the two subspaces.

The last point to be made here is that for ``weak" interactions, weak in the sense that the nonlinear terms in $p_{\alpha a}$ are negligible, the action of $\partial_{\alpha a}$ on a quaternion valued $A\sim \psi_\mu(k_\mu=1)$ single particle state can be calculated.  Define the $\mathbb H$-valued operator 
$\partial_{\alpha a}\to \partial=\partial_0{\bf 1}+\partial_1{\bf i}+\partial_2{\bf j}+\partial_3{\bf k}=\partial_0{\bf 1}+\nabla$, where $\partial_i=\partial/\partial x_i$, such that 
\begin{align*}
\partial A=&{}\partial(A_0{\bf1}+A_1{\bf i}+A_2{\bf j}+A_3{\bf k})=(\partial_0{\bf 1}+\nabla)(A_0{\bf 1}+{\bf A})\\
=&{}(\partial_0 A_0-\nabla\cdot{\bf A}){\bf 1}+(\partial_0{\bf A}+\nabla A_0)+\nabla\times{\bf A}\\
=&{}f{\bf 1}-{\bf E}+{\bf B}
\end{align*}
where notation has been borrowed from $\mathbb R^3$ vector calculus.  The standard interpretation of electric and magnetic fields as derivatives of the vector potential has been made. (The ${\bf E,B}$ fields have to be handled with standard cartesian coordinates to derive Maxwell's equations from this point.)  In the present setting the identity component, $f$, should not be identified with gauge freedom.  As the operators $\partial_{\alpha a}$ act on $\psi_a\in V$ by $\sum_a \partial_{\alpha a}\psi_a$ they sum the ${\bf E,B}$ fields, which corresponds to a macroscopic field acting on a single particle.  Conversely, the conjugate acts on a single $\psi_\alpha\in v$ to react back on $V$.  Recovering something that looks like the electromagnetic field is a self-consistency test.  The Grassmannian has been claimed to convey interactions between subsets, and this derivative recovers one such interaction.  But it also demonstrates that the electromagnetic vector potential is a term in the representation space $\Psi_N$.   The next section will elaborate on other aspects of this changing interpretation of interactions.

\section*{Interpretation of Coordinates}

\subsection*{Action of $G$}
We began by considering a module over a set of coordinates $x$, as these have now been shown to be all that is required to define both the curvature two-forms of the components of the flag manifold and the infinitesmal generators of the Lie algebra (from the Grassmannian).  We did not have to invoke dimensions outside the four-dimensional spacetime manifold in which our primitive points are embedded to obtain the curvature tensors.  We have also seen the left action of $g\in Sp(N)$ on these coordinates: $gxH\to yH$, by linear fractional transformations. In the discussion of eq. (\ref{Hrep}) it was pointed out that subsets of points that are independent of one another are described by the representation $H$, but this is nothing other than a reduced representation of $G$ in eq. (\ref{Grep}).  The same point regarding irreducibility was made in the discussion of curvature following eq. (\ref{MCB}).  Systems that interact with one another are related by an irreducible representation of $G$.  Owing to the coset structure of flag manifolds, $g\in G$ acts on coordinates to the left and the subgroup $H$ acts on the right.  The consequences of this for functions now has to be specified.   

A representation $A_g$ of $g\in G$ acts by\cite{Simon} 
\[
A_g\Psi(x)=\Psi(g^{-1}x),
\]
where $\Psi(x)$ is now a vector in the representation space of $G$ (the subscript $N$ is implicit in the context).  

Geodesics in the group are of the form $\exp(t\mathfrak g), \mathfrak g \in \mathfrak{sp}(N)$, \cite{Price, Simon} where $t\in \mathbb R$ is a universal time coordinate.  The $x_0$ component of each of the quaternions in the flag manifold is a cyclic time-like variable.  One must resist the urge to embed the coset coordinates in an external Euclidean space.  The coordinates define relations between the points in the manifold, and hence relations between the components of the vector bundle $\Psi_N$, independent of any other geometry.  A discussion of the history of ideas about relational \emph{vs.} absolute space can be found in ref. \cite{SEP}.  

\subsection*{Action of $H$}
Having specified the left action of $G$, the action of $H$ remains to be quantified.  We want to define $\Psi_N$ so that it depends only on the coordinates in the flag manifold, and a way to do this is to average over the fiber. A construction from induced representation theory\cite{Folland} is appropriate.  Define 
\[
\Psi(x)=\int \sigma(h)\varphi(xh)dh
\]
to be this average of $\varphi$, where $dh$ is the normalized Haar measure on $h=\bigotimes_\mu h_\mu$ and $\sigma(h)$ is a representation of $h$.  The function $\varphi(xh)$ is a mapping from the group $G\sim Sp(N)$ to our Hilbert space (with dimension appropriate for the dimension of $\sigma$).  The action of $\eta\in h$ on the right of $x$ is 
 \[
\Psi(x\eta)=\int \sigma(h)\varphi(x\eta  h)dh=\int \sigma(\eta^{-1}h)\varphi(xh)dh=\sigma(\eta^*)\Psi(x)
\]
since the normalized Haar measure is invariant to $h\to \eta h$.  Within the $\mu$-th subspace the function $\psi_\mu(x)$ lives in a Hilbert space that is invariant to the action of a representation $\sigma(h_\mu):\psi_\mu(x)\to \sigma(h_\mu)\psi_\mu(x)$.  

The consequences of the action of $H$ \emph{vis-a-vis} bipartite graphs is that one sees the interactions within nodes of a given color only as linear combinations.  To see the interactions amongst the monochromatic nodes in the $\mu$-th subspace in the same way that one sees the interactions between subspaces as described by the action of $G$ on $x$, the $\mu$-th subspace has to be decomposed into single particle states, meaning that $Sp(k)\to Sp(k)/Sp(1)^k$.

 \subsection*{Quaternions and Special Relativity}
A review of well-known facts on the relation between the algebra of quaternions and special relativity is required to set up a mapping between the two.  We begin with the two standard representations of quaternions.  In the first, the basis elements $({\bf 1,i,j,k})$ of the quaternion $x=\{x_0{\bf 1}+x_1{\bf i}+x_2{\bf j}+x_3{\bf k}|x_i\in \mathbb R\}$ are interpreted as abstract algebraic objects in the ring $\mathbb H$ that are subject to the usual rules: ${\bf i}^2={\bf j}^2={\bf k}^2={\bf ijk}=-{\bf 1}$.  The alternative representation makes use of the $M(2,\mathbb C)$ basis of matrix elements over $2\times2$ complex matrices as defined in eq. (\ref{M2C}). We will use both representations in the sequel, and where the context is not clear the basis will be specified as $\mathbb H$ or $M(2,\mathbb C)$.  

The product of two quaternions, $a,b\in\mathbb H$, is 
\begin{equation}\label{prod}
ab=(a_0b_0-{\bf a}\cdot {\bf b}){\bf 1}+a_0{\bf b}+b_0{\bf a}+{\bf a}\times{\bf b}
\end{equation}
where ${\bf a}\cdot {\bf b}=a_1b_1+a_2b_2+a_3b_3$ and ${\bf a}\times{\bf b}=(a_2b_3-a_3b_2){\bf i}+(a_3b_1-a_1b_3){\bf j}+(a_1b_2-a_2b_1){\bf k}$, again borrowing symbols from $\mathbb R^3$ vector calculus for the dot and cross product.  The conjugate of a quaternion $x=x_0{\bf 1}+{\bf x}$ is $\bar x=x_0{\bf 1}-{\bf x}$.  Using the product in eq. (\ref {prod}) it is easy to show that $\overline{ab}=\bar b\bar a$.  Quaternions form a ring, whereas a vector is a module.  One consequence of this is that conjugation by a unit quaternion in $\mathbb H$ is equivalent to the action of $SO(3)$ as a rotation in $\mathbb R^3$.  

Now we turn to the other part of the relation in the title of this section.  Special relativity establishes an equivalence class of frames based on the principle that unit speed (suitably defined) is invariant.  Spacetime coordinates, $(ct,x,y,z)\sim(x_0,x_1,x_2,x_3)$, can be used to construct a quaternion, and the obvious mapping takes $ct\to x_0$. In support of this assignment one notes that the identity component of a quaternion commutes with conjugation by a unit quaternion  (which rotates the spatial components).  To entwine the space and time coordinates the group (acting by conjugation) has to be expanded.  This is done by going over to the $M(2,\mathbb C)$ representation, and noting that $\textrm{det}(q)=|q|$ for $q\in M(2,\mathbb C)$ is the same as $\|q\|^2$ for $q\in \mathbb H$.  The action of $L\in SL(2,\mathbb C)$ by $L:q\to LqL^*$, where $L^*$ is the transpose conjugate of $L$ accomplishes the entwining.  

The group $SL(2,\mathbb C)$ is a manifold of six real dimensions.  Let a matrix in the group be parameterized by $L=\rho\Lambda\tau$, where $\rho\in SU(2)\sim S^3, \tau\in U(2)/U(1)$, and the boost is
\[
\Lambda=\left[\begin{matrix} 
      \lambda & 0 \\
      0 & \lambda^{-1}\\
   \end{matrix}\right], \quad \lambda\in \mathbb R^+.
\]
This parameterization of $L$ is the well-known polar decomposition.  A unit quaternion $u\in SU(2)$ may be parameterized by 
\[
u=\left[\begin{matrix} 
      e^{i\alpha/2} & 0\\
      0 & e^{-i\alpha/2}\\
   \end{matrix}\right]\left[\begin{matrix} 
      \cos(\beta/2) & \sin(\beta/2) \\
      - \sin(\beta/2)& \cos(\beta/2)\\
   \end{matrix}\right]\left[\begin{matrix} 
      e^{i\gamma/2} & 0\\
      0 &  e^{-i\gamma/2}\\
   \end{matrix}\right].
\]
In constructing the polar decomposition of $L$ it is seen that one of the $\textrm{diag}(e^{i\phi/2},e^{-i\phi/2})$ terms in, say the right ($\tau$) side of the diagonal matrix, commutes with the diagonal and is absorbed by $\rho$.  This decomposition accounts for the $3+1+2=6$ real dimensions of a general matrix in $SL(2,\mathbb C)$.  In acting on $x$, the $\rho$ and $\tau$ unitary matrices simply rotate the ``vector" part of the quaternion. The $\Lambda$ piece of $L$ couples the time $x_0\sim ct$ coordinate with just one space coordinate since
\[
x_1=\Lambda \hat x\Lambda:=\left[\begin{matrix} 
      \lambda & 0 \\
      0 & \lambda^{-1}\\
   \end{matrix}\right]\left[\begin{matrix} 
      \zeta_1 & \zeta_2 \\
      -\bar\zeta_2 & \bar\zeta_1\\
   \end{matrix}\right]\left[\begin{matrix} 
      \lambda & 0 \\
      0 & \lambda^{-1}\\
   \end{matrix}\right]=\left[\begin{matrix} 
      \lambda^2\zeta_1 & \zeta_2 \\
      -\bar\zeta_2 & \lambda^{-2}\bar\zeta_1\\
   \end{matrix}\right]
\]
where $\hat x=\tau x\tau^*$.  This is the essential reason that special relativity is silent on rotating frames -- the boost applies to  \emph{only one} space dimension.  (Subsequent rotation of the space frame by $\rho$ simply changes its direction.)  Velocity $v$ is not only the tangent space to $\mathbb R^3$, it can also be interpreted as a projective 3-space, the projection being defined by $iv=(\zeta_1-\bar\zeta_1)/(\zeta_1+\bar\zeta_1)=ix_3/x_0$ (followed by an arbitrary rotation).  This transforms by linear fractional transformations to $i\hat v=(\lambda^2\zeta_1-\lambda^{-2}\bar\zeta_1)/(\lambda^2\zeta_1+\lambda^{-2}\bar\zeta_1)=(S+ivC)/(C+ivS)$ in the boosted coordinates $\hat x$.  Here $C=(\lambda^2+\lambda^{-2})/2$ and $S=(\lambda^2-\lambda^{-2})/2$ are clearly hyperbolic functions since $\lambda$ is real.

The first point to be made about the transformation 
\begin{equation}\label{speed}
i\hat v=(S+ivC)/(C+ivS)
\end{equation}
 is that $v=\pm i$ are two fixed points of the transformation.  The origin $v=0$, is mapped to the imaginary axis $v_1=-iS/C$.  To make the speed a real quantity, either $x_0$ or $x_3$ has to be pure imaginary.  The former choice coincides with a Wick rotation.  The latter choice is equivalent to switching from the $M(2,\mathbb C)$ basis to the Pauli basis.  The two choices are equivalent modulo $i=\sqrt{-1}$.  Here the choice $x_3\to -ix_3, iv\to v$ will be made in eq. (\ref {speed}), so that $\hat v=(S+Cv)/(C+Sv)$.  Now $v=0$ is mapped to $\hat v=S/C$, resulting in the definition of the boost as $(\lambda^2-\lambda^{-2})/2=C=\cosh(2\omega)=1/\sqrt{1-\hat v^2}$.  The fixed points of the map are now at $v=\pm 1$, signifying that the speed of light is the same in the boosted frame as in the original frame.  This is succintly stated: The fixed points at $v=\pm 1$ are invariant to the action of value of $\Lambda$, and hence to the action of $SL(2,\mathbb C)/\pm 1$.  

The second point to be made about the $SL(2,\mathbb C)$ transformation is that the well-known group isomorphism, $SO(1,3)\sim SL(2,\mathbb C)/\pm 1$, allows one to perform the operations above in the $\mathbb R^4$ vector space with coordinates $X=(ct,x,y,z)$,  where $\hat L \in SO(1,3)$ acts by the usual matrix multiplication $\hat L:X\to X\hat L$. The third observation about the relativistic tranformation is that the fixed points of the transformation, $v=\pm 1$, translate into a fixed boundary $c^2t^2-x^2-y^2-z^2=0$ in spacetime.  Of course, all this has been well-known for more than a century.  A review of these facts sets up a particular point of view that is important in the next section.   

\subsection*{The Projection}
A single term in our flag manifold is a quaternion $q_{\mu b}\to x$ with $x\bar x=\|x\|^2$.  Stated differently, $x\bar x=x^2_0+x^2_1+x^2_2+x^2_3=x^{2}_0 + {\bf xx}' =b^2$ defines a three sphere with radius $b>0$. 
Now consider also the hyperbola $y_0^2 - {\bf yy}' = a^2$.  Both $\bf y$ and $\bf x$ are rows consisting of the components of 3-dimensional real vectors.  The family of hyperbolic surfaces, for various values of $a$, has a boundary at the light cone, $a=0$.  The hyperbola and sphere both can be projected into $B^3$, the 3-dimensional ball, by the remarkably similar projections 
\begin{equation}\label{projmap}
{\bf y}/(|y_0|+a) = {\bf x}/(|x_0|+b). 
\end{equation}
On the left
\begin{equation}\label{hyp}
[{\bf y}/(|y_0|+a)][{\bf y}/(|y_0|+a)]'=(|y_0|^2-a^2)/(|y_0|+a)^2=(|y_0|-a)/(|y_0|+a)\le 1.
\end{equation}
This projects both the $y_0 > 0$ (centered at $y_0=-a$) and $y_0<0$ (centered at $y_0=+a$) branches into the ball.   On the sphere side of eq. (\ref{projmap}) we have
\begin{equation}\label{sph}
[{\bf x}/(|x_0|+b)][{\bf x}/(|x_0|+b)]'=(b^2-|x_0|^2)/(b+|x_0|)^2=(b-|x_0|)/(b+|x_0|)\le 1.
\end{equation}
This is a sterographic projection of the northern/southern hemisphere centered at the south/north pole.  Both the hyperbola and sphere require two coordinate charts to cover them.  The ball $B^3$ is a velocity space.  The single quaternion in the Yang-Mills field strength is also handled by the projection in eq. (\ref{projmap}).  However, in the $S^4$ representation the ${\bf x}$ in eq. (\ref{sph}) is four dimensional, which projects to an AdS space. To remedy this there is a stereographic projection of $S^4$ into the quaternions\cite{Gorm}.  These several different representations of four-dimensional spacetime geometry clearly relate to one another through these projections.  

This projective equivalence shows that one may convert a (spherical) quaternion in the Grassmannian into the corresponding (hyperbolic) ``Pauli pseudo-quaternion", so the two are not so different.  However, this only makes sense one term at a time -- one must not project the Grassmannian as a whole, as that would defeat the group structure.  This is not a defect on the Grassmannian side.  The Grassmannian is a many-body construction, whereas relativity is strictly valid only as a pair-wise relation between frames. 

The light cone boundary, $a=0$ in eq. (\ref{hyp}), is the boundary of the ball, which is a two sphere.  This corresponds to $x_0=0$ in eq. (\ref{sph}).  One may interpret a photon as an interaction with vanishing identity component, suggesting that the identity component of a quaternion in the Grassmannian is related to mass.

\section*{Acknowledgment}

The author is grateful for several helpful discussions with Profs. John Sullivan and Gerald Folland.

\end{document}